\documentclass[journal]{IEEEtran}
\usepackage{multirow}
\usepackage{diagbox}
\usepackage{amssymb}
\usepackage{amsmath}
\usepackage{graphicx}
\usepackage{cite}
\usepackage{citesort}
\usepackage{subfigure}
\usepackage{graphicx,epstopdf}
\usepackage{epsfig}	
\usepackage{cite,graphicx,amsmath,amssymb}
\usepackage{comment}
\usepackage{algorithm}
\usepackage{algorithmic}
\usepackage{caption}
\usepackage{amssymb}
\usepackage{amsmath}
\usepackage{cite}
\usepackage{url}
\usepackage{xcolor}
\usepackage{cite,graphicx,amsmath,amssymb}
\usepackage{subfigure}
\usepackage{citesort}
\usepackage{fancyhdr}
\usepackage{mdwmath}
\usepackage{mdwtab}
\usepackage{caption}
\usepackage{amsthm}
\usepackage{lipsum}

\newtheorem{theorem}{Theorem}

\newtheorem{lemma}{Lemma}

\newtheorem{corollary}{Corollary}

\makeatletter
\def\ScaleIfNeeded{%
\ifdim\Gin@nat@width>\linewidth \linewidth \else \Gin@nat@width
\fi } \makeatother

\begin{document}

\title{\Huge{STAR-RIS Aided Secure MIMO Communication Systems }}
\author{Xiequn Dong, Zesong Fei,~\IEEEmembership{Senior~Member,~IEEE}, Xinyi Wang,~\IEEEmembership{Member,~IEEE}, Meng Hua,~\IEEEmembership{Member,~IEEE}, Qingqing Wu,~\IEEEmembership{Senior~Member,~IEEE}

\thanks{Xiequn Dong, Zesong Fei, and Xinyi Wang are with the School of Information and Electronics, Beijing Institute of Technology, Beijing 100081, China (e-mail: dongxiequn@bit.edu.cn, feizesong@bit.edu.cn, bit\_wangxy@163.com.).

M. Hua is with the Department of Electrical and Electronic Engineering, Imperial College London, London SW7 2AZ, UK (e-mail: m.hua@imperial.ac.uk).

Qingqing Wu is with the Department of Electronic Engineering, Shanghai Jiao Tong University, Shanghai 200240, China (e-mail: qingqingwu@sjtu.edu.cn).

}
}

\maketitle

\begin{abstract}
This paper investigates simultaneous transmission and reflection reconfigurable intelligent surface (STAR-RIS) aided physical layer security (PLS) in multiple-input multiple-output (MIMO) systems, where the base station (BS) transmits secrecy information with the aid of STAR-RIS against multiple eavesdroppers equipped with multiple antennas. We aim to maximize the secrecy rate by jointly optimizing the active beamforming at the BS and passive beamforming at the STAR-RIS, subject to the hardware constraint for STAR-RIS. To handle the coupling variables, a minimum mean-square error (MMSE) based alternating optimization (AO) algorithm is applied. In particular, the amplitudes and phases of STAR-RIS are divided into two blocks to simplify the algorithm design. Besides, by applying the Majorization-Minimization (MM) method, we derive a closed-form expression of the STAR-RIS's phase shifts. Numerical results show that the proposed scheme significantly outperforms various benchmark schemes, especially as the number of STAR-RIS elements increases.


\end{abstract}
\section{Introduction}
The 6th-generation (6G) wireless communication network has put forward higher requirements for throughput as well as security. Reconfigurable intelligent surface (RIS), a meta-surface with massive passive reflecting elements, has evolved as a revolutionary technology for 6G wireless communications \cite{9903378,10422712}. Through changing the phase shifts and/or amplitudes of incident signals, RIS has the capacity to intelligently modify wireless propagation environment \cite{8910627,Hua,Wang,Dong,10284917,Li2}. In addition, different types of RIS also expand more communication scenarios, such as active RIS \cite{Chen}.
Furthermore, RIS is a flexible technology. In combination with other technologies, such as integrated sensing and communication technology, RIS expands the sensing function and further improves the quality of communication \cite{Meng1,Meng2}.  The over-the-horizon sensing capabilities provided by RIS can help enable non-line-of-sight eavesdroppers to perceive, thereby enhancing the security of communications. However, the conventional RIS can only serve the users within its front half-space due to the hardware constraint.

To address the limitation of conventional RIS, simultaneous transmitting and reflecting RIS (STAR-RIS) has been introduced as a supplement \cite{9690478,9774942,9754364,Li1}. Although the STAR-RIS significantly widens the service coverage, it also increases the risk of information leakage to the suspicious eavesdroppers due to the full coverage of 360 degrees. To address this challenge, researchers are studying how to improve the secrecy performance with the assistance of STAR-RIS in \cite{10138694,9734006,10005206,9739715}. To restrain the internal eavesdropping in STAR-RIS-aided communication systems, the authors in \cite{10138694} proposed a secure transmission design via joint optimization of active and passive beamforming with the aim of maximizing the secrecy rate. Besides, in \cite{9734006}, the authors studied a secrecy rate maximization problem in a multiple-input multiple-output (MIMO) system and proposed a minimum mean-square error (MMSE) method to handle the non-convex quadratically constrained quadratic program (QCQP) problem. Furthermore, in \cite{10005206}, the authors considered the scenario where separate eavesdroppers are located in reflection and transmission regions, respectively, and proposed a block coordinate descent (BCD) based optimization framework to maximize the worst secrecy rate under imperfect channel state information (CSI) of eavesdroppers. A penalty-based successive convex approximation (SCA) method based on S-Procedure was used to solve the non-convex problem. Besides, in \cite{9739715}, artificial noise (AN) was embedded into signals to enhance the secrecy performance. However, the aforementioned schemes \cite{10138694,9734006,10005206,9739715} cannot be directly applied in the MIMO case. For security scenarios, the 360-degree coverage of STAR-RIS brings more risks of eavesdropping. Motivated by the challenge of the increased risk of eavesdropping due to the full coverage character of STAR-RIS in this MIMO case, we propose a novel secrecy joint active and passive beamforming framework to handle the eavesdropping in STAR-RIS aided MIMO secrecy communication systems.

In this paper, we aim to maximize the secrecy rate of STAR-RIS assisted secure MIMO communication networks that one eavesdropper and one legitimate user in both the transmission region and reflection region by jointly optimizing the active and passive beamforming. The STAR-RIS is working in energy splitting (ES) mode. Compared with the other two modes, ES mode has the advantages of low implementation complexity than time splitting mode and high degree of freedom \cite{9690478}. The formulated problem is non-convex due to the highly coupled variables. To handle this problem, we divide the problem into three sub-problems under the framework of alternating optimization (AO). In particular, we use the MMSE method to convert the active beamforming design sub-problem into a QCQP form and further transform it into a convex form. For passive beamforming optimization, we divide it into phase and amplitude optimization sub-problems. For phase optimization, we adopt Majorization-Minimization (MM) method to derive a closed-form solution. For amplitude optimization, we transform the original sub-problem into a convex subproblem. Simulation results demonstrate that compared with existing schemes, the proposed algorithm is able to achieve $10\%$ improvement in terms of secrecy rate, and the performance improvement becomes more significant with the increase of transmission energy and the number of STAR-RIS elements.
\section{System Model and Problem Formulation}
\begin{figure}[t!]
\centering
\includegraphics[width=3.5in,  height=2.4in]{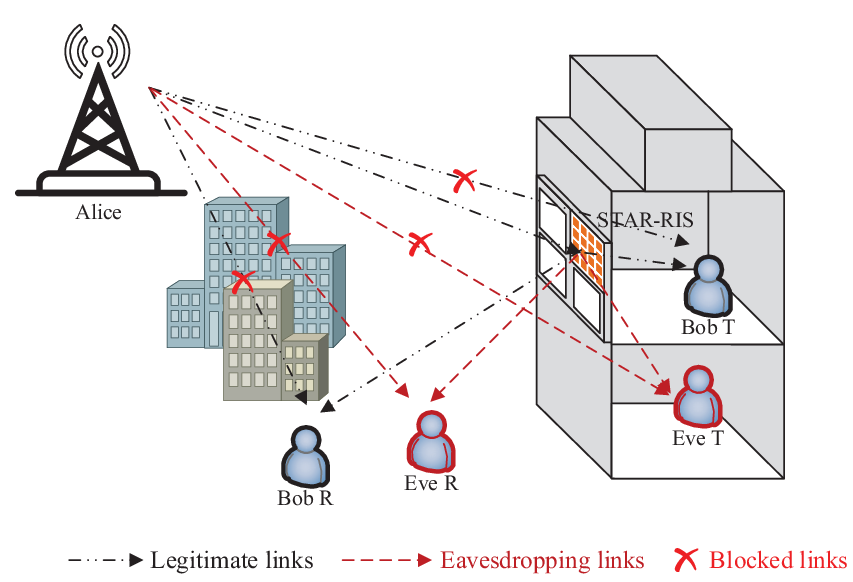}
\caption{An illustration of the STAR-RIS-aided secure MIMO communication system.}
\label{Fig. 1}
\end{figure}
We consider a STAR-RIS assisted secure MIMO communication system as shown in Fig. \ref{Fig. 1}, which consists of an $N$-antenna base station (BS), an $L$-element STAR-RIS, two $Z$-antenna legitimate users, and two $M$-antenna eavesdroppers.\footnote{Note that the case with multiple users and multiple eavesdroppers, the algorithm proposed in this paper still applies since it has no impact on the final expression of the secrecy rate.
} The STAR-RIS is an $L_xL_y$-element rectangular plane, with $L_x$ and $L_y$ representing the horizontal and vertical number of elements at the STAR-RIS panel, respectively. Thus, the total number of  elements is given by $L = L_xL_y$. There exist one eavesdropper and one legitimate user in both transmission region and reflection region. Moreover, we assume that the eavesdroppers are communication nodes with advanced computing capability but low security level in this system. As a result, perfect CSI of eavesdroppers are assumed to be available for BS. As shown in Fig. 1, since there is no direct links, the BS is required to transmit signals to users with the help of STAR-RIS. The transmit signals can be expressed as
\begin{align}\label{1}
{\mathbf{x}} = {\mathbf{w}_r}{s_r} + {\mathbf{w}_{t}}{s_{t}},
\end{align}
where $s_r$ and $s_t$ denote signals transmitted to Bob R and Bob T, respectively. ${{\mathbf{w}_r} \in {\mathbb{C}^{N \times 1}}}$ and ${{\mathbf{w}_t} \in {\mathbb{C}^{N \times 1}}}$ are the beamforming for ${s_r}$ and ${s_t}$, with $s_r$ and $s_t$ both assumed to follow a circularly symmetric complex Gaussian distribution, i.e., $\mathbb{E}\left[s_k^2\right] = 1$, ${k \in \left\{ {t,r} \right\}}$.

Therefore, the signals received at the users and the eavesdroppers can be expressed as
\begin{align}\label{2}
{\mathbf{y}_{k,b}} = {\mathbf{T}_k}{\mathbf{\Theta} _k}\mathbf{H}{\mathbf{x}} + {\mathbf{n}_{k,b}},
\end{align}
and
\begin{align}\label{3}
{\mathbf{y}_{k,e}} = {\mathbf{G}_k}{\mathbf{\Theta} _k}\mathbf{H}{\mathbf{x}} + {\mathbf{n}_{k,e}},
\end{align}
respectively, where ${\mathbf{\Theta} _r} = {\rm{diag}}\left( {\sqrt {{a _{1,r}}} {e^{j{\theta _{1,r}}}}, \ldots \sqrt {{a _{L,r}}} {e^{j{\theta _{L,r}}}}} \right) \in \mathbb{C}{^{L \times L}}$ and ${\mathbf{\Theta} _t} = {\rm{diag}}\left( {\sqrt {{a _{1,t}}} {e^{j{\theta _{1,t}}}}, \ldots \sqrt {{a _{L,t}}} {e^{j{\theta _{L,t}}}}} \right) \in \mathbb{C}{^{L \times L}}$ denote the reflection coefficient matrix and transmission coefficient matrix of STAR-RIS, respectively. ${\mathbf{H}} \in {{\mathbb{C}}^{L \times N}}$, ${\mathbf{T}_k} \in {{\mathbb{C}}^{Z \times L}}$, and ${\mathbf{G}_k} \in {{\mathbb{C}}^{M \times L}}$ denote the baseband equivalent channels of the BS to the STAR-RIS, the STAR-RIS to the eavesdroppers, and the STAR-RIS to the user links, respectively. We consider all links to be Rician fading links. Besides, ${\mathbf{n}_{k,b}} \sim {\cal C}{\cal N}\left( {\mathbf{0},{\sigma_{k,b} ^2\mathbf{I}}} \right)$ and ${\mathbf{n}_{k,e}} \sim {\cal C}{\cal N}\left( {\mathbf{0},{\sigma_{k,e} ^2\mathbf{I}}} \right)$ denote the complex additive white Gaussian noise at the users and the eavesdroppers, respectively.

The received achievable rates at users and eavesdroppers can be given by
\begin{align}\label{4}
{R_{k,b}} = \log \det \Big( {\mathbf{I} + {{{ {{\mathbf{T}_k}{\mathbf{\Theta} _k}\mathbf{H}{\mathbf{w}_k}{\mathbf{w}_k^H}\mathbf{H}^H{{\mathbf{\Theta}} _k^H}{\mathbf{T}_k}^H} }}}} \Big.\\\nonumber
\Big. {{\left( {{{\mathbf{T}_{k}}{\mathbf{\Theta} _{k}}\mathbf{H}{\mathbf{w}_{k'}}{\mathbf{w}_{k'}^H}\mathbf{H}^H{{\mathbf{\Theta}} _{k}^H}{\mathbf{T}_{k}}^H} + {\sigma_{k,b} ^2}\mathbf{I}} \right)}^{ - 1}} \Big),
\end{align}
and
\begin{align}\label{5}
{R_{k,e}} = \log \det \Big( {\mathbf{I} + {{{ {{\mathbf{G}_k}{\mathbf{\Theta} _k}\mathbf{H}{\mathbf{w}_k}{\mathbf{w}_k^H}\mathbf{H}^H{\mathbf{\Theta} _k^H}{\mathbf{G}_k^H}} }}}} \Big.\\\nonumber
\Big.  {{\left( {{{\mathbf{G}_k}{\mathbf{\Theta} _k}\mathbf{H}{\mathbf{w}_{k'}}{\mathbf{w}_{k'}^H}\mathbf{H}^H{\mathbf{\Theta} _k^H}{\mathbf{G}_k^H}}+ {\sigma_{k,e} ^2}\mathbf{I}} \right)}^{ - 1}}\Big),
\end{align}
respectively.
The secrecy rate $R_{k,s}$ between the BS and legitimate user $k$ can be given by \cite{9734006}
\begin{align}\label{6}
{R_{k,s}} = \left[ {0,{R_{k,b}} - {R_{k,e}}} \right]^{+}.
\end{align}

In this paper, we aim to maximize the secrecy rate by jointly optimizing active and passive beamforming at the STAR-RIS assisted MIMO wireless communication systems. Accordingly, the problem can be formulated as
\begin{align}\label{7}
\mathop {{\rm{maximize}}}\limits_{{\pmb\Theta _k},{\mathbf{w}_k}} &\sum\limits_k {{R_{k,s}}} \\\nonumber
\rm{s.t.}\;&{\rm{ C1:}}\sum\limits_k {\mathbf{w}_k^H{\mathbf{w}_k}} \le P,\\\nonumber
&{\rm{C}}2: 0 \le {a _{l,r}} \le 1,0 \le {a _{l,t}} \le 1, {a _{l,r}} + {a _{l,t}} \le 1,\\\nonumber &\;\;\;\;\;\;\;1 \le l \le L,\\\nonumber
&{\rm{C}}3:0 \le {\theta _{l,k}} \le 2\pi (k \in \left\{ {r,t} \right\}),\\\nonumber
\end{align}
where $P$ is the maximum transmit power at BS. $\rm{C1}$ represents the power constraint to be satisfied by beamforming. $\rm{C2}$ represents the amplitude constraint to be satisfied for each element of STAR-RIS. $\rm{C3}$ represents the phase constraint to be satisfied by each element of STAR-RIS. Problem (\ref{7}) is non-convex due to the coupling of beamforming and the phase shifts. Hence, in the following section, we adopt an AO scheme based on MMSE and MM algorithm to optimize the active and passive beamforming, iteratively.
\section{Proposed Algorithm}
In this section, we propose an AO algorithm to solve problem (\ref{7}).
\subsection{MMSE-based Beamforming Optimization Algorithm }
For any given $\mathbf{\Theta}$, problem (\ref{7}) is still non-convex due to its objective function. To handle this non-convex term, we convert it to the following form
\begin{align}\label{9}
{R_{k,s}} = &\underbrace {\log \left| {\mathbf{I} +  {{{{\mathbf{\tilde H}}}_k}{{\mathbf{w}}_k}{\mathbf{w}}_k^H{\mathbf{\tilde H}}_k^H} }  {{{\left( {{\sigma_{k,b} ^2}{\mathbf{I}} + {{{\mathbf{\tilde H}}}_k}{{\mathbf{w}}_{k'}}{\mathbf{w}}_{k'}^H{\mathbf{\tilde H}}_k^H} \right)}^{ - 1}}} \right|}_{g_{k,1}}\\\nonumber
&\underbrace { + \log \left| {{{ {\mathbf{\tilde G}_k{\mathbf{w}_{k'}}{\mathbf{w}^H_{k'}}\mathbf{\tilde G}_k^H} }}{{\left( {{\sigma_{k,e} ^2}\mathbf{I}} \right)}^{ - 1}} + \mathbf{I}} \right|}_{{g_{k,2}}}\\\nonumber
&\underbrace { - \log \left| \mathbf{I} +  {\sigma_{k,e} ^{ - 2}}{{ {\mathbf{\tilde G}_k\left({\mathbf{w}_k}{\mathbf{w}_k}^H + {\mathbf{w}_{k'}}{\mathbf{w}_{k'}^H}\right)\mathbf{\tilde G}_k^H}}}  \right|}_{{g_{k,3}}},
\end{align}
where $\mathbf{\tilde H}_k = {\mathbf{T}_k}{\mathbf{\Theta} _k}\mathbf{H}$, and $\mathbf{\tilde G}_k = {\mathbf{G}_k}{\mathbf{\Theta} _k}\mathbf{H}$. To deal with the non-convex term in (\ref{9}), we firstly transform $g_{k,1}$ into an equivalent form based on the MMSE method \cite{9279253}. Specifically, $g_{k,1}$ can be viewed as the data rate for a hypothetical communication system where the user estimates the desired signal $s_k$ with an estimator ${\mathbf{u}_{k,1}} \in {\mathbb{C}^{Z \times 1}}$, the estimated signal is given by
\begin{align}\label{10}
\hat s_k = \mathbf{u}_{k,1}^H{\mathbf{y}_{k,b}}.
\end{align}
Therefore, the MSE expression can be given by
\begin{align}\label{11}
{E_{k,1}} = &\mathbb{E}\left\{ {{{\left( {{{\hat s}_k} - {s_k}} \right)}^2}} \right\}\\\nonumber
 = &\left( {\mathbf{u}_{k,1}^H{\mathbf{\tilde H}_k}{\mathbf{w}_k} - 1} \right){\left( {\mathbf{u}_{k,1}^H{\mathbf{\tilde H}_k}{\mathbf{w}_k} - 1} \right)^H}\\\nonumber
 &+ \mathbf{u}_{k,1}^H\left( {{\mathbf{\tilde H}_k}{\mathbf{w}_{k'}}\mathbf{w}_{k'}^H\mathbf{\tilde H}_k^H + \sigma _{k,b}^2{\mathbf{I}}} \right){\mathbf{u}_{k,1}}.
\end{align}
Through introducing additional variables $W_{k,1}$ and $\mathbf{u}_{k,1}\in {\mathbb{C}^{Z \times 1}}$, we can convert $g_{k,1}$ into the following expression
\begin{align}\label{12}
{{\hat g}_{k,1}} = &\mathop {\max }\limits_{{W_{k,1}} > 0,{\mathbf{u}_{k,1}}} \log \left( {{W_{k,1}}} \right)\\\nonumber&- {W_{k,1}}{E_{k,1}}\left( {{\mathbf{u}_{k,1}},{\mathbf{w}_k},{\mathbf{w}_{k'}}} \right) + 1,
\end{align}
where ${W_{k,1}} = E_{k,1}^{ - 1}\left( {{{\mathbf{u}}_{k,1}},{\mathbf{w}_k},{\mathbf{w}_{k'}}} \right)$ and ${\mathbf{u}_{k,1}} = {\left( {{\sigma_{k,b} ^2}\mathbf{I} + {\mathbf{\tilde H}_k}\left( {{\mathbf{w}_k}\mathbf{w}_k^H + {\mathbf{w}_{k'}}\mathbf{w}_{k'}^H} \right)\mathbf{\tilde H}_k^H} \right)^{ - 1}}{\mathbf{H}_k}{\mathbf{w}_k}$. Similarly, by introducing $W_{k,2}$, $\mathbf{u}_{k,2}$ and $\mathbf{W}_{k,3}$, $g_{k,2}$ and $g_{k,3}$ can be converted as
\begin{align}\label{13}
{{\hat g}_{k,2}} = &\mathop {\max }\limits_{{{W}_{k,2}} > 0,{\mathbf{u}_{k,2}}} \log \left( {{W_{k,2}}} \right) - {W_{k,2}}{E_{k,2}}\left( {{\mathbf{u}_{k,2}},{\mathbf{w}_{k'}}} \right) + N,
\end{align}
and
\begin{align}\label{14}
{{\hat g}_{k,3}} = &\mathop {\max }\limits_{{{\mathbf{W}}_{k,3}} > 0} \log \left| {\mathbf{W}_{k,3}} \right| - {\rm{Tr}}\left( {{\mathbf{W}_{k,3}}{\mathbf{E}_{k,3}}\left( {{\mathbf{w}_k},{\mathbf{w}_{k'}}} \right)} \right) + Z,
\end{align}
respectively, where ${E_2}\left( {{\mathbf{u}_{k,2}},{\mathbf{w}_{k'}}} \right) = \left( {\mathbf{u}_{k,2}^H{\mathbf{\tilde G}_k}{\mathbf{w}_{k'}} - 1} \right){\left( {\mathbf{u}_{k,2}^H{\mathbf{\tilde G}_k}{\mathbf{w}_{k'}} - 1} \right)^H} + \mathbf{u}_{k,2}^H\left( {\sigma _{k,e}^2\mathbf{I}} \right){\mathbf{u}_{k,2}}$, ${{{W}}_{k,2}} = E_{k,2}^{ - 1}\left( {{\mathbf{u}_{k,2}},{\mathbf{w}_{k'}}} \right)$, ${\mathbf{u}_{k,2}} = {\left( {{\sigma_{k,e} ^2}\mathbf{I} + {\mathbf{\tilde G}_k}{\mathbf{w}_{k'}}\mathbf{w}_{k'}^H\mathbf{\tilde G}_k^H} \right)^{ - 1}}{\mathbf{\tilde G}_k}{\mathbf{w}_{k'}}$, ${\mathbf{E}_{k,3}}\left( {{\mathbf{w}_k},{\mathbf{w}_{k'}}} \right) = \mathbf{I} + \sigma _{k,e}^{ - 2}{{\mathbf{\tilde G}}_k}\left( {{\mathbf{w}_{k'}}\mathbf{w}_{k'}^H + {\mathbf{w}_k}\mathbf{w}_k^H} \right)\mathbf{\tilde G}_k^H$, ${{\mathbf{W}}_{k,3}} = \mathbf{E}_{k,3}^{ - 1}\left( {{\mathbf{w}_k},{\mathbf{w}_{k'}}} \right)$.

Thereafter, (\ref{9}) can be expressed as
 \begin{align}\label{16}
{R_{k,s}} =  &- \mathbf{w}_k^H{\mathbf{A}_k}{\mathbf{w}_k} - \mathbf{w}_{k'}^H{\mathbf{B}_k}{\mathbf{w}_{k'}}\\\nonumber
 &+ {W_1}\left( {\mathbf{u}_{k,1}^H{\mathbf{\tilde H}_k}{\mathbf{w}_k} + \mathbf{w}_k^H\mathbf{\tilde H}_k^H{\mathbf{u}_{k,1}}} \right) \\\nonumber
 &+ {W_2}\left( {\mathbf{u}_{k,2}^H{\mathbf{\tilde G}_k}{\mathbf{w}_{k'}} + \mathbf{w}_{k'}^H\mathbf{\tilde G}_k^H{\mathbf{u}_{k,2}}} \right) + d_k\\\nonumber
 \triangleq & - f\left( {{\mathbf{w}_k},{\mathbf{w}_{k'}}} \right),
\end{align}
where ${\mathbf{A}_k} = \mathbf{\tilde H}_k^H{\mathbf{u}_{k,1}}{W_{k,1}}\mathbf{u}_{k,1}^H{{\mathbf{\tilde H}}_k} + \sigma _{k,e}^{ - 2}\mathbf{\tilde G}_k^H{\mathbf{W}_{k,3}}{{\mathbf{\tilde G}}_k}$. ${\mathbf{B}_k} = \mathbf{\tilde H}_k^H{\mathbf{u}_{k,1}}{W_1}\mathbf{u}_{k,1}^H{{\mathbf{\tilde H}}_k} + \mathbf{\tilde G}_k^H{\mathbf{u}_{k,2}}{W_{k,2}}\mathbf{u}_{k,2}^H{{\mathbf{\tilde G}}_k} + \sigma _{k,e}^{ - 2}\mathbf{\tilde G}_k^H{\mathbf{W}_{k,3}}{{\mathbf{\tilde G}}_k}$. $d_k = \log \left( {{W_{k,1}}} \right) + 1 + \log \left( {{W_{k,2}}} \right) + N + \log \left| {{\mathbf{W}_{k,3}}} \right| + Z
 - {W_{k,1}}\left( {1 + \mathbf{u}_{k,1}^H\left( {\sigma _{k,b}^2\mathbf{I}} \right){\mathbf{u}_{k,1}}} \right) - {W_{k,2}}\left( {1 + \mathbf{u}_{k,2}^H\left( {\sigma _{k,e}^2\mathbf{I}} \right){\mathbf{u}_{k,2}}} \right) - {\rm{Tr}}\left( {{\mathbf{W}_{k,3}}} \right)$.
Therefore, for given reflection and transmission coefficient matrices, we can convert (\ref{7}) to the following convex form
 \begin{align}\label{17}
&\mathop {{\rm{minimize}}}\limits_{{\mathbf{w}_r},{\mathbf{w}_{t}}} \sum\limits_k {f\left( {{\mathbf{w}_k},{\mathbf{w}_{k'}}} \right)} \\\nonumber
&\rm{s.t.}\;\mathbf{w}_r^H{\mathbf{w}_r} + \mathbf{w}_{t}^H{\mathbf{w}_{t}} \le P,
\end{align}
which can be solved via CVX \cite{CVX}.
\renewcommand{\algorithmicrequire}{\textbf{Input:}}
\renewcommand{\algorithmicensure}{\textbf{Output:}}

\subsection{MM-based Algorithm}
For given $\mathbf{w}_r$ and $\mathbf{w}_t$, the optimization problem (\ref{7}) can be expressed as
\begin{align}\label{18}
&\mathop {{\rm{maximize}}}\limits_{{\pmb\Theta _k}}\sum\limits_k {{R_{k,s}}} \\\nonumber
&\rm{s.t.} \;\rm{C2-C3}.
\end{align}
Based on (\ref{17}) and (\ref{18}), by defining $\mathbf{W}_k = \mathbf{w}_k{\mathbf{w}_k^H}$, we can reformulate the objective function as (\ref{19}),
 \begin{figure*}
 	\centering
 	\begin{align}\label{19}
g\left( {{\pmb\Theta _k}} \right) = &{W_{k,1}}\left( {\mathbf{u}_{k,1}^H{{\mathbf{\tilde H}}_k}{\mathbf{W}_k}\mathbf{\tilde H}_k^H{\mathbf{u}_{k,1}} - \mathbf{u}_{k,1}^H{{\mathbf{\tilde H}}_k}{\mathbf{w}_k} - \mathbf{w}_k^H\mathbf{\tilde H}_k^H{\mathbf{u}_{k,1}}} \right) \\\nonumber
&+ {W_{k,2}}\left( {\mathbf{u}_{k,2}^H{\mathbf{{\tilde G}}_k}{\mathbf{W}_{k'}}\mathbf{\tilde G}_k^H{\mathbf{u}_{k,1}} - \mathbf{u}_{k,2}^H{\mathbf{{\tilde G}}_k}{\mathbf{w}_{k'}} - \mathbf{w}_{k'}^H\mathbf{\tilde G}_k^H{\mathbf{u}_{k,2}}} \right)
 + \sigma _{k,e}^2{\rm{Tr}}\left( {{\mathbf{W}_{k,3}}\left( {{{\mathbf{\tilde G}}_k}\left( {{\mathbf{W}_k} + {\mathbf{W}_{k'}}} \right)\mathbf{\tilde G}_k^H} \right)} \right) - {d_k}\\\nonumber
 = &{\rm{Tr}}\left( {\pmb\Theta _k^H{\mathbf{X}_{k,1}}{\pmb\Theta _k}{\mathbf{Y}_{k,1}}} \right) - {\rm{Tr}}\left( {{\pmb\Theta _k}{\mathbf{Z}_{k,1}}} \right) - {\rm{Tr}}\left[ {{{\left( {{\pmb\Theta _k}{\mathbf{Z}_{k,1}}} \right)}^H}} \right] \\\nonumber
 &+ {\rm{Tr}}\left( {\pmb\Theta _k^H{\mathbf{X}_{k,2}}{\pmb\Theta _k}{\mathbf{Y}_{k,2}}} \right) - {\rm{Tr}}\left( {{\pmb\Theta _k}{\mathbf{Z}_{k,2}}} \right) - {\rm{Tr}}\left[ {{{\left( {{\pmb\Theta _k}{\mathbf{Z}_{k,2}}} \right)}^H}} \right] + {\rm{Tr}}\left( {\pmb\Theta _k^H{\mathbf{X}_{k,3}}{\pmb\Theta _k}{\mathbf{Y}_{k,3}}} \right) - {d_k},
 	\end{align}
\hrulefill
\vspace*{4pt}
 \end{figure*}
where ${\mathbf{X}_{k,1}} = \mathbf{T}_k^H{\mathbf{u}_{k,1}}{W_{k,1}}\mathbf{u}_{k,1}^H{\mathbf{T}_k}$, ${\mathbf{Y}_{k,1}} = \mathbf{H}{\mathbf{W}_k}{\mathbf{H}^H}$, ${\mathbf{Z}_{k,1}} = \mathbf{H}{\mathbf{w}_k}W_{k,1}{\mathbf{u}_{k,1}^H}{\mathbf{T}_k}$, ${\mathbf{X}_{k,2}} = \mathbf{G}_k^H{\mathbf{u}_{k,2}}{W_{k,2}}\mathbf{u}_{k,2}^H{\mathbf{G}_k}$, ${\mathbf{Y}_{k,2}} = \mathbf{H}{\mathbf{W}_{k'}}{\mathbf{H}^H}$, ${\mathbf{Z}_{k,2}} = \mathbf{H}{\mathbf{w}_{k'}}W_{k,2}{\mathbf{u}_{k,2}^H}\mathbf{T}_k^H$, ${\mathbf{X}_{k,3}} = {\sigma_{k,e} ^{ - 2}}\mathbf{G}_k^H{\mathbf{W}_3}{\mathbf{G}_k}$, ${\mathbf{Y}_{k,3}} = \mathbf{H}({\mathbf{W}_r}+{\mathbf{W}_{t}}){\mathbf{H}^H}$.

Based on the above reformulation, by vectorizing the RIS matrix ${\pmb{\theta} _k} = {\left[ {\sqrt {{a _{1,k}}} {e^{j{\theta _{1,k}}}}, \ldots \sqrt {{a _{L,k}}} {e^{j{\theta _{L,k}}}}} \right]^T} \in L \times 1$, we can convert (\ref{19}) as
\begin{align}\label{20}
g\left( {{\pmb{\theta} _k}} \right) = \;&\pmb{\theta} _k^H{\mathbf{\Gamma} _{k,1}}\pmb\theta_k  + \pmb\theta _k^H{\mathbf{\Gamma} _{k,2}}\pmb\theta_k  + \pmb\theta _k^H{\mathbf{\Gamma} _{k,3}}\pmb\theta_k \\\nonumber
 &- \left[ {\left( {{\mathbf{z}_{k,1}} + {\mathbf{z}_{k,2}}} \right)}^H {\pmb\theta _k}\right] - \left[ {\pmb\theta _k^H\left( {{\mathbf{z}_{k,1}} + {\mathbf{z}_{k,2}}} \right)} \right]- d_k,
\end{align}
where ${\mathbf{\Gamma} _{k,1}} = {\mathbf{X}_{k,1}} \odot \mathbf{Y}_{k,1}^T$, ${\mathbf{\Gamma} _{k,2}} = {\mathbf{X}_{k,2}} \odot \mathbf{Y}_{k,2}^T$, ${\mathbf{\Gamma} _{k,3}} = {\mathbf{X}_{k,3}} \odot \mathbf{Y}_{k,3}^T$, with $\odot$ representing the Hadamard product operator. $\mathbf{z}_{k,1}$ and $\mathbf{z}_{k,2}$ are the vectors composed of the diagonal elements of matrix $\mathbf{Z}_{k,1}$ and $\mathbf{Z}_{k,2}$. Hence, the problem (\ref{18}) can be expressed as
\begin{align}\label{21}
&\mathop{\rm{minimize}}\limits_{{\pmb\theta _k}}\sum\limits_k {g\left( {{\pmb{\theta} _k}} \right)} \\\nonumber
&{\rm{s.t.\;}}{\rm{diag}}\left( {{\pmb{\theta} _k}\pmb{\theta} _k^H + {\pmb{\theta}_{k'}}\pmb{\theta}_{k'}^H} \right)\preceq{\mathbf{1}_L},
\end{align}
where ${\mathbf{1}_L} = \left[ {1, \ldots ,1} \right] \in {\mathbb{C}^{1 \times L}}$. Furthermore, we divide the optimization of the RIS part into amplitude and phase optimization. For the optimization of phase shifts, since $\mathbf{\Gamma} _{k,1}$, $\mathbf{\Gamma} _{k,2}$ and $\mathbf{\Gamma} _{k,3}$ in (\ref{20}) are semidefinite, we rewrite (\ref{20}) as
\begin{align}\label{22}
g\left( {{\pmb{\theta} _k}} \right) = \;&\pmb{\theta} _k^H{\mathbf{\Gamma}_k}\pmb\theta_{k} - \left[ {\left( {{\mathbf{z}_{k,1}} + {\mathbf{z}_{k,2}}} \right)}^H {\pmb\theta _k}\right] \\\nonumber
- &\left[ {\pmb\theta _k^H\left( {{\mathbf{z}_{k,1}} + {\mathbf{z}_{k,2}}} \right)} \right]- d_k,
\end{align}
where $\mathbf{\Gamma}_k = \mathbf{\Gamma}_{k,1} + \mathbf{\Gamma}_{k,2} + \mathbf{\Gamma}_{k,3}$.
\begin{lemma}
According to \cite{9279253}, the upper-bounded function $g( {\left. {{\pmb\theta _k}} \right|{{\pmb{\tilde \theta} }_k}} )$ for $g\left( {{\pmb\theta _k}} \right)$ can be expressed as
\begin{align}\label{23}
g\left( {\left. {{\pmb\theta _k}} \right|{\pmb{\tilde \theta }_k}} \right) = &{\lambda _{k,\max }}{\pmb{\theta}} _k^H{\pmb\theta _k} - 2{\mathop{\rm Re}\nolimits} \left\{ {\pmb\theta _k^H\left( {{\lambda _{k,\max }}{{\mathbf{I}}_L} - {\pmb\Gamma _k}} \right){\pmb{\tilde \theta }_k}} \right\}\\\nonumber
 + &\pmb{\tilde \theta} _k^H\left( {{\lambda _{k,\max }}{{\mathbf{I}}_L} - {\pmb\Gamma _k}} \right){\pmb{\tilde \theta }_k} - \left[ {\left( {{\mathbf{z}_{k,1}} + {\mathbf{z}_{k,2}}} \right)}^H {\pmb\theta _k}\right]\\\nonumber
 - &\left[ {\pmb\theta _k^H\left( {{\mathbf{z}_{k,1}} + {\mathbf{z}_{k,2}}} \right)} \right] - {d_k},
\end{align}
where $\lambda_{\max}$ is the maximum eigenvalue of $\pmb\Gamma_k$.
\end{lemma}Hence, the subproblem can be given as
\begin{align}\label{24}
&\mathop{\rm{minimize}}\limits_{{\pmb\theta _k}}\sum\limits_k {g\left( {\left. {{\pmb\theta _k}} \right|{\pmb{\tilde \theta }_k}} \right)} \\\nonumber
&{\rm{s.t.\;}}{\rm{ }}{\rm{diag}}\left( {{\pmb{\theta} _k}\pmb{\theta} _k^H + {\pmb{\theta}_{k'}}\pmb{\theta}_{k'}^H} \right)\preceq{\mathbf{1}_L}.
\end{align}
Since ${\pmb{\theta}} _k^H{\pmb\theta _k} = L_k(L_k = a_{1,k}+...+a_{L,k})$, we can rewrite the objective function as
\begin{align}\label{25}
g\left( {\left. {{\pmb\theta _k}} \right|{{\pmb{\tilde \theta} }_k}} \right) = {\lambda _{k,\max }}L_k + 2{\mathop{\rm Re}\nolimits} \left\{ {{\pmb\theta _k^H}{\mathbf{q}_k}} \right\},
\end{align}
where ${\mathbf{q}_k} = \left[ {\left( {{\pmb\Gamma _k} - {\lambda _{k,\max }}{{\mathbf{I}}_L}} \right){\pmb{\tilde \theta }_k} - \left( {{\mathbf{z}_{k,1}} + {\mathbf{z}_{k,2}}} \right)} \right]$. Hence, the optimal phase $\pmb\varphi_k$ can be given by
\begin{align}\label{26}
\pmb\varphi_k  = {e^{ j\arg \left( {{{\mathbf{-q}}_k}} \right)}}.
\end{align}

Afterwards, we optimize the amplitude of the reflection and transmission coefficients. Based on (\ref{26}), we can rewrite $\pmb\theta_k$ as
\begin{align}\label{27}
{\pmb\theta _k} = \pmb\Phi_k\mathbf{a}_k,
\end{align}
where $\pmb\Phi_k=\rm{diag}(\pmb\varphi_k)$, $\mathbf{a}_k=[\sqrt{a_{1,k}},...,\sqrt{a_{L,k}}]^T$. Therefore, the objective function of problem (\ref{21}) can be rewritten as 
\begin{align}\label{28}
g\left( {{\mathbf{a} _k}} \right) = \;&\mathbf{a} _k^H{\mathbf{\tilde\Gamma} _{k,1}}\mathbf{a}_k  + \mathbf{a} _k^H{\mathbf{\tilde\Gamma} _{k,2}}\mathbf{a}_k  + \mathbf{a} _k^H{\mathbf{\tilde\Gamma} _{k,3}}\mathbf{a}_k - d_k \\\nonumber
 &- \left[ {\left( {{\mathbf{z}_{k,1}} + {\mathbf{z}_{k,2}}} \right)}^H \pmb\Phi_k{\mathbf{a} _k}\right] - \left[ {\mathbf{a} _k^H\pmb\Phi_k^H\left( {{\mathbf{z}_{k,1}} + {\mathbf{z}_{k,2}}} \right)} \right],
\end{align}
where $\mathbf{\tilde\Gamma} _{k,1} = \pmb\Phi_k^H\mathbf{\Gamma} _{k,1}\pmb\Phi_k$, $\mathbf{\tilde\Gamma} _{k,2} = \pmb\Phi_k^H\mathbf{\Gamma} _{k,2}\pmb\Phi_k$, $\mathbf{\tilde\Gamma} _{k,3} = \pmb\Phi_k^H\mathbf{\Gamma} _{k,3}\pmb\Phi_k$. Hence, amplitude optimization problem is converted to the following form.
\begin{align}\label{29}
&\mathop{\rm{minimize}}\limits_{{\mathbf{a} _k}}\sum\limits_k {g\left( {{\mathbf{a} _k}} \right)} \\\nonumber
&{\rm{s.t.\;}}{\rm{diag}}\left( {{\mathbf{a} }_k\mathbf{a} _k^H + {\mathbf{a} _{k'}}\mathbf{a} _{k'}^H} \right)\preceq{\mathbf{1}_L}.
\end{align}
This problem is a typical convex QCQP problem and can be solved by convex optimization tools such as CVX.

\subsection{Overall Algorithm and Complexity Analysis}
By alternatively solving (\ref{17}), (\ref{24}), and (\ref{29}) until the objective function converges, the original problem (\ref{7}) can be solved. The detailed steps are summarized in Algorithm 1. The complexity of beamforming optimization and phase optimization are ${\cal O}\left({4Z^3 + 8 + 2M^3}\right)$ and ${\cal O}\left({L^3+L^2}\right)$ \cite{9279253}. The complexity of amplitude optimization is ${\cal O}\left({L^3}\right)$ \cite{9734006}. Therefore, the total complexity of Algorithm 1 is ${\cal O}\left(K_{iter}\left( {4Z^3 + 8 + 2M^3 + 2L^3 + L^2} \right)\right)$, where $K_{iter}$ represents the total number of iterations required by Algorithm 1 to converge.
\begin{algorithm}\label{Algorithm 1}
	\caption{Overall Algorithm for Solving Problem (\ref{7})}
	\label{alg3}
	\begin{algorithmic}[1]
        \STATE Initialize:\;$\pmb{\varphi}^{(0)}_r$, $\pmb{\varphi}^{(0)}_t$, $\mathbf{a}^{(0)}_r$, $\mathbf{a}^{(0)}_t$, $\mathbf{w}_{r}^{(0)}$, $\mathbf{w}_{t}^{(0)}$, and $\epsilon_1$.
        \STATE Set $n=0$.
		\REPEAT
		\STATE Solve (\ref{17}) for given $\pmb{\varphi}^{(n)}_r, \pmb{\varphi}^{(n)}_t, \mathbf{a}^{(n)}_r, \mathbf{a}^{(n)}_t, \mathbf{w}_{r}^{(n)}$, and $\mathbf{w}_{t}^{(n)}$. Obtain the optimal beamforming $\mathbf{w}_r$ and $\mathbf{w}_t$. Update $\mathbf{w}_{r}^{(n+1)}=\mathbf{w}_r$ and $\mathbf{w}_{t}^{(n+1)}=\mathbf{w}_t$.
        \STATE Solve (\ref{24}) for given $\pmb{\varphi}^{(n)}_r$, $\pmb{\varphi}^{(n)}_t$, $\mathbf{a}^{(n)}_r$, $\mathbf{a}^{(n)}_t$, $\mathbf{w}_{r}^{(n+1)}$, and $\mathbf{w}_{t}^{(n+1)}$. Obtain the optimal phase shifts $\pmb{\varphi}_r$ and $\pmb{\varphi}_t$. Update $\pmb{\varphi}^{(n+1)}_r=\pmb{\varphi}_r$ and $\pmb{\varphi}^{(n+1)}_r=\pmb{\varphi}_t$.
        \STATE Solve (\ref{29}) for given $\pmb{\varphi}^{(n+1)}_r$, $\pmb{\varphi}^{(n+1)}_t$, $\mathbf{a}^{(n)}_r$, $\mathbf{a}^{(n)}_t$, $\mathbf{w}_{r}^{(n+1)}$, and $\mathbf{w}_{t}^{(n+1)}$. Obtain the optimal amplitude $\mathbf{a}_r$ and $\mathbf{a}_t$. Update $\mathbf{a}^{(n+1)}_r=\mathbf{a}_r$ and $\mathbf{a}^{(n+1)}_t=\mathbf{a}_t$.
		\UNTIL convergence.
		\ENSURE $\pmb{\varphi}_r$, $\pmb{\varphi}_t$, $\mathbf{a}_r$, $\mathbf{a}_t$, $\mathbf{w}_{r}$ and $\mathbf{w}_{t}$.
	\end{algorithmic}
\end{algorithm}

\section{Numerical results}
In this section, we verify the effectiveness of the proposed algorithm via numerical results. The 3D coordinates of STAR-RIS, BS, Bob R, Eve R, Bob T, and Eve T are $[0,0,30]\rm{m}$, $[100,0,30]\rm{m}$, $[120,20,0]\rm{m}$, $[150,150,0]\rm{m}$, $[-120,0,30]\rm{m}$, $[-120,50,60]\rm{m}$. The path loss exponents of all the links are set as $\alpha = 2.2$, the Rician factor is set as $K=3$, and the path loss at the reference distance of 1 meter is set as $\rho_{0}=-30$ $\rm{dB}$ \cite{9423667}. The distance between two adjacent elements/antennas $d = \lambda/2$. The noise power is $\sigma_{k,b}^{2} = \sigma_{k,e}^{2} = -90$ $\rm{dBm}$. In order to show the performance gain brought by the proposed scheme in STAR-RIS assisted secure communication systems, we also consider the following benchmark schemes: 1) MMSE-SDR scheme \cite{9279253}: the semidefinite relaxation (SDR) method is applied in the optimization of STAR-RIS. In the part of phase optimization, we set ${\pmb\Phi _k} = {\pmb\theta _k}\pmb\theta _k^H$. Hence, we can rewrite (18) as
\begin{align}\label{1}
{g_k}\left( {{\pmb\Phi _k}} \right){\rm{ = Tr}}\left( {{\pmb\Pi _k}{\pmb\Phi _k}} \right) - {d_k},
\end{align}
where ${\Pi _k} = \left[ {\begin{array}{*{20}{c}}
{{\pmb\Gamma _k}}&{ - \left( {{\mathbf{z}_{k,1}} + {\mathbf{z}_{k,2}}} \right)}\\
{ - \left( {\mathbf{z}_{k,1}^H + \mathbf{z}_{k,2}^H} \right)}&0
\end{array}} \right]$. Hence, the problem (16) in the manuscript can be expressed as
\begin{align}\label{2}
&\mathop {{\rm{minimize}}}\limits_{{\pmb\Phi _k}} {\rm{  }}\sum\limits_k {g\left( {{\pmb\Phi _k}} \right)} \\\nonumber
&{\rm{s}}{\rm{.t}}{\rm{. diag}}\left( {{\pmb\Phi _r}} \right) + {\rm{diag}}\left( {{\pmb\Phi _t}} \right) = \mathbf{1}_L,\\\nonumber
&{\pmb\Phi _k} \succeq 0,\rm{rank}\left( {{\pmb\Phi _k}} \right) = 1,k \in \left\{ {r,t} \right\}.
\end{align}
However, the existence of the rank-one constraint makes the subproblem non-convex, so we choose to ignore the rank-one constraint in the process of solving and use Gaussian randomization to restore a rank-one solution. Moreover, the resulting $\pmb\Phi _k$ is decomposed by EVD
\begin{align}\label{100}
\pmb\Phi_k  = \mathbf{U}_k\pmb\Sigma_k {\mathbf{U}_k^H},
\end{align}
based on (\ref{100}), we can obtain the vector $\mathbf{v}_k = \mathbf{U}_k{\pmb\Sigma_k ^{1/2}}\mathbf{r}_k$, where $\mathbf{r} \in {\cal{CN}}\left( {\mathbf{0},{\mathbf{I}_{L + 1}}} \right)$. Next, among all the $\mathbf{v}$ obtained by different $\mathbf{r}$, the $\mathbf{v}$ that can maximize the objective function is selected. Finally, $\mathbf{\bar v} = {e^{j\arg \left( {{{\left[ {\frac{\mathbf{v}}{{{\mathbf{v}_{L + 1}}}}} \right]}_{\left( {1:L} \right)}}} \right)}}$ is used to satisfy the constant modulus constraint.; 2) MMSE-QCQP scheme \cite{9734006}: the subproblem of STAR-RIS's phase shifts optimization is transformed as a QCQP problem. In this part, we can expressed the phase subproblem as
\begin{align}\label{3}
&\mathop{\rm{minimize}}\limits_{{\pmb\theta _k}}\sum\limits_k {g\left( {{\pmb{\theta} _k}} \right)} \\\nonumber
&{\rm{s.t.\;}}{\rm{diag}}\left( {{\pmb{\theta}} _k\pmb{\theta} _k^H + {\pmb{\theta}}_{k'}\pmb{\theta}_{k'}^H} \right)={\mathbf{1}_L}.
\end{align}
However, due to the constant modulus constraint, this subproblem is non-convex. To make the problem tractable, we loosen the constant mode constraint as
\begin{align}\label{4}
{\rm{diag}}\left( {{\pmb\theta _k}\pmb\theta _k^H + {\pmb\theta _{k'}}\pmb\theta _{k'}^H} \right) \preceq \mathbf{1}_L
\end{align}
Based on (\ref{3}) and (\ref{4}), the subproblem is convex and can be solved by CVX toolbox; 3) MRT scheme: the subproblem of beamforming optimization is solved by maximum ratio transmission (MRT) method.

\begin{figure}[t!]
\centering
\includegraphics[width=3in,  height=2.4in]{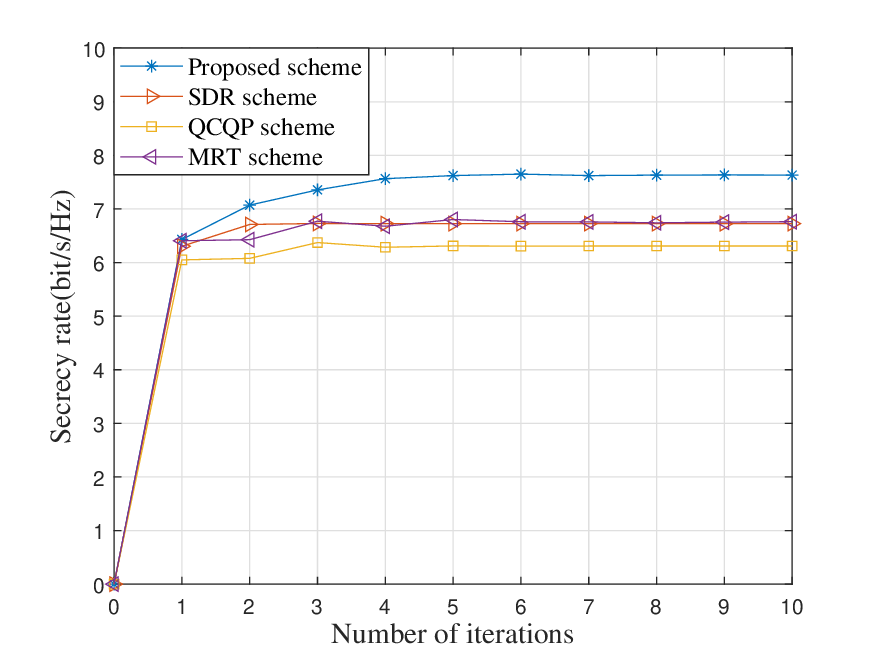}
\caption{Convergence of the proposed algorithm.}
\label{Fig. 2}
\end{figure}

Fig. \ref{Fig. 2} shows the convergence of the proposed scheme as well as the other benchmark schemes. The maximum transmit power of the BS is 30 $\rm{dBm}$. The numbers of antennas at the BS, users, and eavesdroppers are both set as 4. Furthermore, the number of RIS elements is 20 $\left( L_x =5, L_y = 4 \right)$. Although it requires more iterations for convergence compared to the MRT scheme, MMSE-SDR scheme and MMSE-QCQP scheme, the proposed scheme is able to achieve the highest secrecy rate, which validates the superiority of the proposed algorithm.

\begin{figure}[t!]
\centering
\includegraphics[width=3in,  height=2.4in]{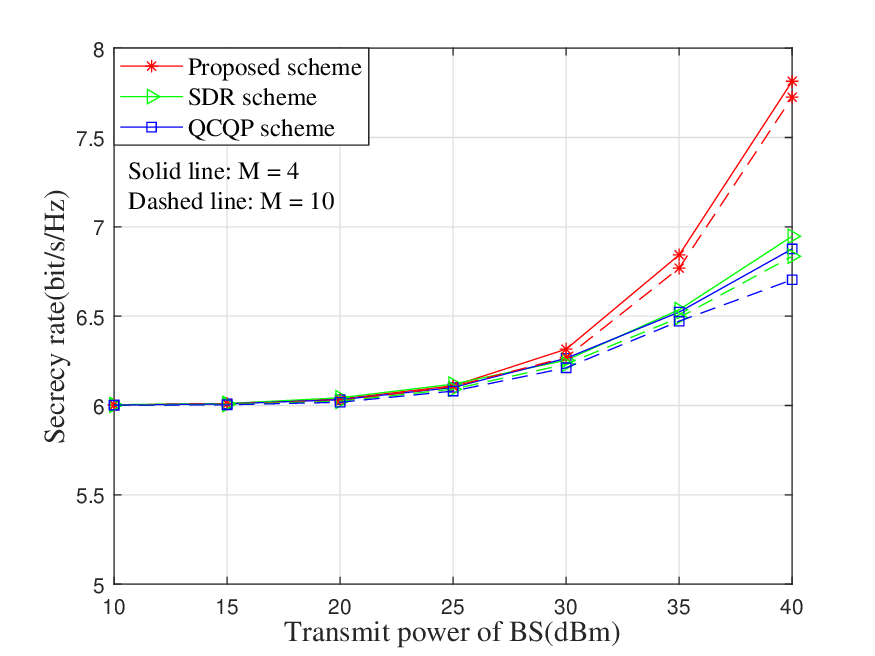}
\caption{Secrecy rate versus the transmit power of BS.}
\label{Fig. 4}
\end{figure}

In Fig. \ref{Fig. 4}, we show the secrecy rate versus the transmit power of BS. As we can see, with the increase of BS transmit power, the secrecy performance of systems with the STAR-RIS are increasing because the transmit power becomes the dominant factor affecting the secrecy rate. Moreover, we can see that when the transmission power is low, the performance of the three schemes is nearly the same, but with the increase of transmit power, the performance of the proposed scheme significantly outperforms the other schemes. This is because the other two comparison schemes loosen the constraints when optimizing the phase shifts, resulting in significant performance loss when recovering the solution to satisfy the original constraints. Furthermore, this affects the performance of the overall optimization algorithm and reduces the benefits that STAR-RIS can bring to the system. Therefore, with the increase of transmit power, the proposed scheme can bring more improvement than the other schemes. Besides, When the number of antennas of the eavesdropper increases, the secrecy rate decreases. This is due to an increase in the antenna gain at the eavesdropper.

\begin{figure}[t!]
\centering
\includegraphics[width=3in,  height=2.4in]{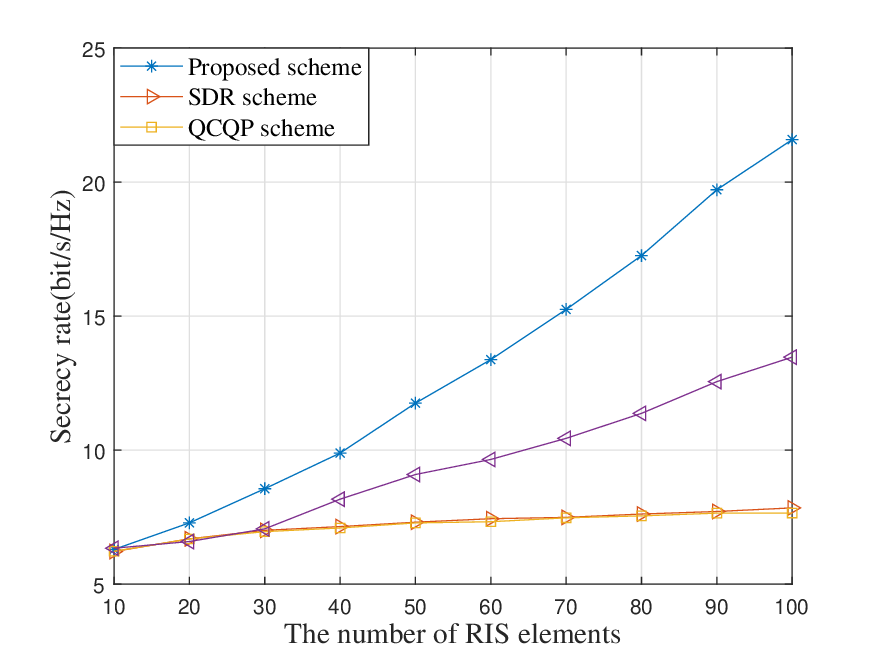}
\caption{Secrecy rate versus the number of STAR-RIS elements.}
\label{Fig. 5}
\end{figure}

In Fig. \ref{Fig. 5}, we show the secrecy rate versus the number of elements at the STAR-RIS. As we can see, with the increase of the number of elements at the STAR-RIS, the secrecy performance of systems with the STAR-RIS is increasing. In addition, with the increase of the STAR-RIS's element numbers, the performance gap between the proposed scheme and other schemes becomes larger. This is because the two comparison schemes loosen the constant modulus constraint, resulting in reduced performance when optimizing phase shifts. The proposed scheme also brings more freedom to the deployment of the number of STAR-RIS elements.

\section{Conclusion}
In this paper, we investigated joint active and passive beamforming, phase shifts, and amplitudes design for STAR-RIS-aided MIMO downlink communication systems to improve secrecy performance. To solve the non-convex problem, an efficient AO algorithm was proposed. In particular, we proposed an MMSE-based optimization algorithm for active beamforing, while for the optimization of the STAR-RIS phase shifts, we divided original problem into phase optimization problem and amplitude optimization problem. In regard of phase optimization, MM algorithm was utilized to obtain closed-form solutions. In amplitude optimization, the QCQP algorithm was used. Simulation results showed that the proposed scheme is effective in improving the secrecy performance.

\bibliographystyle{IEEEtran}
\bibliography{mybib}

\end{document}